\documentclass[a4paper]{article}
\usepackage{algorithmic}

\usepackage{algorithm,xcolor}
\usepackage[percent]{overpic}
\usepackage{hyperref}
\usepackage{INTERSPEECH2022}
\usepackage{soul}

\hypersetup{
    colorlinks=true,
    linkcolor=blue,
    filecolor=red,      
    urlcolor=blue,
    }

\title{Zero-Shot Voice Conditioning for Denoising Diffusion TTS Models}
\name{Alon Levkovitch$^1$, Eliya Nachmani$^{1,2}$, Lior Wolf$^1$}

\address{
  Tel Aviv University}
  \address{
  $^1$Tel Aviv University \quad   $^2$Facebook AI Research}
\email{}

\begin{document}

\maketitle
\begin{abstract}
We present a novel way of conditioning a pretrained denoising diffusion speech model to produce speech in the voice of
a novel person unseen during training. The method requires a
short ($\sim 3$ seconds) sample from the target person, and generation is steered at inference time, without any training
steps. At the heart of the method lies a sampling process that
combines the estimation of the denoising model with
a low-pass version of the new speaker’s sample. The objective and
subjective evaluations show that our sampling method can
generate a voice similar to that of the target speaker in terms of frequency, with an accuracy comparable to state-of-the-art methods,
and without training.
\end{abstract}

\vspace{-0.3cm}
\section{Introduction}

Denoising Diffusion Probabilistic Models (DDPM) have emerged as leading generative models in speech and other domains~\cite{ho2022cascaded,chen2020wavegrad, dhariwal2021diffusion}. When applied to spectrograms, they generate state-of-the-art speech in applications such as Text-To-Speech (TTS)~\cite{popov2021grad} and neural vocoders \cite{kong2021diffwave,chen2020wavegrad}.

The iterative nature of the denoising generation scheme creates an opportunity to modify the process. The Iterative Latent Variable Refinement (ILVR) method~\cite{choi2021ilvr} does so for images by steering the generated image toward a low-resolution template. This way, the output image acquires some of the properties of the target template.

In this work, we explore the possibility of employing a similar steering mechanism, with voice generation based on spectrograms. We find that this is an accessible way of converting the output voice of a pretrained DDPM TTS model, which does not require any finetuning and uses only a short sample of the target voice. 

Through objective evaluation, our results show that our method is able to generate speech that approximates the reference speaker in the frequency domain. The subjective evaluation shows that the generated speech is similar to that of the reference speaker and its naturalness is comparable to state-of-the-art methods. 

\section{Background}
%\vspace{-0.5cm}
\noindent{\bf Voice conversion methods\quad}
  Voice conversion methods can be divided into three main groups that are each based on different methods: (i) speaker embedding conversion, (ii) conversion using fine-tuning on short audio clips, and (iii) style encoding. 
  
  Speaker embedding methods, such as DGC-VECTOR \cite{xiao2022dgc}, AutoVC \cite{qian2019autovc}, SEVC \cite{tan2021zero}, {YourTTS \cite{casanova2021yourtts}} IZSVC \cite{nercessian2020improved} and VoiceLoop \cite{taigman2017voiceloop},  use a generation process conditioned on speaker embedding. During training, these embeddings are calculated for the training set. For voice conversion, these methods encode unseen speech into a new speaker embedding and use it to generate audio. 
  
  Methods that use fine-tuning, such as VoiceGrad \cite{kameoka2020voicegrad} and TTS Skins \cite{polyak2019tts} use a pre-trained model, which they fine-tune on a short sample from a new speaker. The model can then generate audio samples based on the new speaker. TTS Skins uses a TTS robot as base voice, converting it into a new speaker. 
  
  Methods that use style encoding, such as StarGANv2-VC \cite{li2021starganv2}, Stargan-vc2 \cite{kaneko2019stargan}, StarGAN-ZSVC \cite{baas2021stargan} and STDR \cite{yuan2021improving}, employ a style encoder in their generation process. The style encoder conditions the generation on the style of a reference. However, for best performance, the encoder and the whole model have to be fine-tuned for the new encoding styles of a new speaker. 
  
\noindent{\bf Score-based Generative Modeling\quad} In score-based generative models, the original data is corrupted with Gaussian noise according to a stochastic differential equation (SDE):
\begin{equation}
    dx = f(x,t)dt  + g(x,t)dw
    \label{eq1}
\end{equation}

 where $w$ is the standard Weiner process, $t$ is a value sampled from a finite time interval $[0,T]$, $f(\cdot. t)$ is the drift coefficient of x and $g(\cdot. t)$ is the diffusion coefficient of x.
 
 An important property of the forward diffusion process is that the reverse process is also a diffusion process. Let $\hat{w}$ denote the reverse time Wiener process. The reverse time SDE~\cite{elliott1985reverse} is:
\begin{multline}
    dx = [f(x,t) - g(x,t)g(x,t)^T\nabla_{x}\log pt(x)]dt\\
    + g(x,t)d\hat{w}
    \label{eq2}
\end{multline}

 Furthermore, Kolmogorov’s forward equation of Eq.\ref{eq1} is equivalent to a more simplified probability flow ordinary differential equation \cite{song2020score}. This observation implies that both equations ultimately describe the same statistical process, albeit in a different form.
\begin{multline}
    dx = f(x,t)dt - \frac{1}{2}\nabla [g(x,t)g(x,t)^T]dt\\ - \frac{1}{2}g(x,t)g(x,t)^T\nabla_{x}\log pt(x)dt
    \label{eq3}
\end{multline}

 Therefore, given a time-dependent model $s_{\theta}(x, t)$ that can approximate the score $\nabla_x \log pt(x)$, we can sample $x_T \sim p_T$ and compute reverse steps according to Eq.\ref{eq3} to obtain $x_0 \sim p_0$ from the original data distribution.
 
\noindent{\bf Grad-TTS\quad} 
~\cite{popov2021grad} is a pre-trained, score-based Text-To-Speech (TTS) model. Grad-TTS uses a text encoder to define a prior distribution $\mu$ and then generates spectrograms from this distribution. In the last step, it uses a vocoder to convert the generated spectrograms into audio. We use this model to demonstrate our method. The forward SDE of Grad-TTS is defined as:
\begin{equation}
    dx_t = \frac{1}{2}\Sigma^{-1}(\mu - x_t)\beta_t dt + \sqrt{\beta_t}dw\,,
    \label{eq4}
\end{equation}
 where $\mu$ is the mean of the prior distribution, $\Sigma$ is the covariance matrix, and $\beta_t$ is the noise schedule. The mean is estimated by a prior text encoder. In Grad-TTS, the covariance is assumed as an identity matrix (we keep this variable in the exposition for the sake of completeness).

 Since the drift coefficient $f(\cdot, t)$ is affine, the transition kernel $q_\theta(x_t | x_0)$ is a Gaussian distribution, whose mean and variance can be obtained in closed form. Let $\gamma_t = \exp(-\int_{0}^{t} \beta_s ds)$ The transition kernel can be written as:
\begin{equation}
    q_{\theta}(x_t | x_0) = \mathcal{N}((\mathbf{I} - \sqrt{\gamma_t})\mu + \sqrt{\gamma_t})x_0, \Sigma (\mathbf{I} - \gamma_t))
    \label{eq5}
\end{equation}

 On an infinite time horizon, where $\lim_{t\to T}\gamma_t = 0$, the transition kernel simply converges to $\mathcal{N}(\mu, \Sigma)$. In other words, the forward process guarantees that diffusion will yield samples from the prior distribution.
 The reverse SDE of Grad-TTS can be derived using Eq.\ref{eq2} and Eq.\ref{eq4}:
\begin{equation}
    dx_t = \frac{1}{2}(\Sigma^{-1}(\mu - x_t)) - \nabla_x \log p_t(x_t))\beta_t dt
    \label{eq6}
\end{equation}

 Therefore, given a score estimator $s_\theta$ and an initial condition from the prior distribution $\mathcal{N}(\mu, \Sigma)$, we can compute Eq.\ref{eq6} to sample from the learned data distribution.

A more in-depth discussion of its components, such as monotonic alignment search~\cite{kim2020glow}, can be found in the original paper by Popov et al.~\cite{popov2021grad}. 

 \begin{figure*}[t]
 \vspace{-0.5cm}

\centering
\begin{overpic}[width=.8\linewidth]{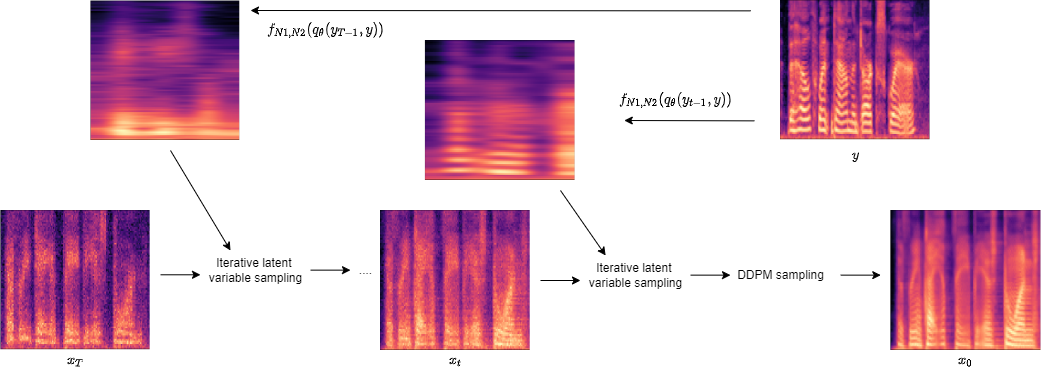}
\end{overpic}
\label{fig1}

\caption{An overview of our method, which applies iterative latent variable sampling to the Grad-TTS model. The method starts with $x_T$, which is noise conditioned on text, refining $x_T$ until it is a clear voice at $x_0$. During the process, the latent variables are mixed with the reference voice $y$ by a low-pass filter $f$ and the forward diffusion process $q_\theta$.}
\vspace{-0.5cm}
\end{figure*}
\noindent{\bf Iterative Latent Variable Refinement} (ILVR)~\cite{choi2021ilvr} is a method for conditioning the generative process of an unconditional DDPM model to generate images that share the high-level semantics of a given reference image. 
 
 The conditional sampling process relies on the conditional distribution $p(x_0|c)$ with the condition $c$:
\begin{gather}
    p_\theta(x_0|c) = \int p_\theta(x_{0:T}|c)dx_{1:T}\\
    p_\theta(x_{0:T}|c) = p(x_T)\prod_{t=1}^Tp_\theta(x_{t-1}|x_t, c)
    \label{eq7}
\end{gather}
 In this process, each transition $p_\theta(x_{t-1}|x_t, c)$ depends on the condition $c$. Often, one may wish to use an unconditionally trained DDPM that represents an unconditional transition $p_\theta (x_{t-1},x_t)$, as in Eq.\ref{eq6}. ILVR provides sample-based conditioning, given an unconditioned transition $p_\theta (x_{t-1},x_t)$, without requiring additional training or models. It does this by refining each unconditional transition with downsampled reference data.
 
 Let $\phi_N(\cdot)$ denote a linear low-pass filtering operation, a downsampling and upsampling operation by a factor of $N$, thus maintaining the spatial dimensions of the data. Given a reference data item $y$, the condition $c$ is to ensure that the downsampled data $\phi_N(x_0)$ of the generated data $x_0$ is equal to $\phi_N(y)$.
 Thus, let $x'_{t-1} \sim p_\theta(x'_{t-1}|x_t)$, $y_{t-1} \sim q_\theta(y_{t-1}|y)$, the latent update step is:
 \begin{equation}
     x_{t-1} \longleftarrow \phi_N(y_{t-1}) + x'_{t-1} - \phi_N(x'_{t-1})
     \label{eq8}
 \end{equation}
 Where $q_\theta$ is the forward diffusion process defined by the unconditional model. This update step forces the condition that the downsampled data be equal and thus forces the unconditional DDPM model to generate samples that are close to the reference in the output space of $\phi_N(\cdot)$.
 
\section{Proposed Method}

 Our method conditions the Grad-TTS DDPM on an audio clip of a reference speaker. This adds a new condition to the Grad-TTS DDPM, which is already conditioned on text and a defined speaker from the model's training. This new conditioning forces the features of the model's output to be similar to that of the reference speaker instead of the speaker the model was trained on. Since we use a low-pass filter, the spoken text of the reference does not need to match the text encoded by $\mu$. 
 
We define a low-pass filter $f_{N_{\rm F}, N_{\rm T}}(\cdot)$ that applies bicubic downsampling and upsampling on spectrograms. Since spectrograms do not have the same scale on the frequency and time axes, we use two scale factors, $N_{\rm F}$ and $N_{\rm T}$. These correspond to frequency and time, respectively. Using our low-pass filter as part of an ILVR sampling process, we steer Grad-TTS DDPM to generate samples similar to the reference under the low-pass operation.
 
 Grad-TTS DDPM can be formulated as a conditioned generation model:
 \begin{equation}
    x_{t-1} = p_\theta(x_{t-1}|x_t, \mu, t, j)\\
     \label{eq9}  
 \end{equation}

 where $\mu$ is the text distribution, $t$ is the time step and $j$ is the speaker embedding. For brevity, let us rename these to condition $c$, such that:
 \begin{equation}
     p_\theta(x_{t-1}|x_t, \mu, t, j) \rightarrow p_\theta(x_{t-1}|x_t, c)
 \end{equation}

 From the definition of the SDE, we can also formulate:
\begin{equation}
\begin{aligned}
     x_{t-1} & = x_T - \sum_{i=t}^{T} dx_i
     &= x_T - \sum_{i=t}^{T} \frac{1}{2}(\Sigma^{-1}(\mu-x_i)) - s_\theta(x_t, c)\beta_i
     \label{eq10}
\end{aligned}
 \end{equation}
 
 In our method, we use ILVR to add a new condition to the model using the low-pass filter, $f(\cdot)=f_{N_{\rm F}, N_{\rm T}}(\cdot)$. We can then formulate the condition as:
 \begin{equation}
     x_{t-1} = p_\theta(x_{t-1}|x_t, c, f(x_0) = f(y))
     \label{eq11} 
 \end{equation} 
  where $y$ is the reference spectrogram and $x_0$ is the spectrogram generated in the last step of the process. We have that 
 \begin{equation}
     x_0 = x_t - \sum_{i=1}^{t} \frac{1}{2}(\Sigma^{-1}(\mu-x_i)) - s_\theta(x_i, c)\beta_i\,.
     \label{eq12}
 \end{equation}
 Plugging this into Eq.\ref{eq11}, and approximating the Markov transitions of the process: 
 \begin{equation}
 \begin{aligned}
    &x_{t-1} = p_\theta(x_{t-1}|x_t, c, f(x_{t-1} - \sum_{i=1}^{t-1} dx_i) = f(y)) \approx \\
    &\mathbb{E}_{q_\theta(y_{t-1}|y)}[p_\theta(x_{t-1}|x_t, c, f(x_{t-1} - \sum_{i=1}^{T} dx_i) = f(y_{t-1} - \sum_{i=1}^{t-1}dy_i))]
    \label{eq13}
 \end{aligned}
 \end{equation}
 Since the forward diffusion and $f$ are linear operators, this can be simplified further, obtaining
 \begin{equation}
 \begin{aligned}
   x_{t-1} &= \mathbb{E}_{q_\theta(y_{t-1}|y)}[p_\theta(x_{t-1}|x_t, c,\\ & \quad\ f(x_{t-1}) = f(y_{t-1})), f(s_\theta(x_{t-1}, c)) = f(s_\theta(y_{t-1}, c))] \\
    &\approx \mathbb{E}_{q_\theta(y_{t-1}|y)}[p_\theta(x_{t-1}|x_t, c, f(x_{t-1}) = f(y_{t-1}))] 
    \label{eq14}
\end{aligned}
\end{equation}
Let $x'_{t-1} = p_\theta(x_{t-1}|x_t, c)$ be the proposal, without our new condition. It is then refined via $f(x_{t-1}) = f(y_{t-1})$. Thus, we obtain:
 \begin{equation}
 \begin{aligned}
 &\mathbb{E}_{q_\theta(y_{t-1}|y)}[p_\theta(x_{t-1}|x_t, c, f(x_{t-1}) = f(y_{t-1}))] = \\
 &\mathbb{E}_{q_\theta(y_{t-1}|y)}[p_\theta(f(y_{t-1}) + \\
 &(\mathbf{}{I} - f) x'_{t-1} |x_t, c,\
 f(x_{t-1}) = f(y_{t-1}))] = \\
 &\mathbb{E}_{q_\theta(y_{t-1}|y)}[p_\theta(x'_{t-1}|x_t,c)]= \\
 &p_\theta(x'_{t-1}|x_t, c) = \\
 &p_\theta(x_{t-1}|x_t, c, f(x_{t-1}) = f(y_{t-1}))
 \end{aligned}
 \label{eq15}
 \end{equation}
 
 From this, we get the ILVR update step: 
 \begin{equation}
     x_{t-1} = f(y_{t-1}) + x_{t-1}' - f(x_{t-1}')\,,
     \label{eq16}
 \end{equation}
 which supports conditioning on a new voice without any additional training.
 
 Since using this update step during the entire sampling process would introduce too much noise, at a pre-defined time step we move from updating with Eq.\ref{eq16} back to using only Eq.\ref{eq6}. We define this time step as a parameter called "ILVR stopping step". Thus, we can formulate the sampling process as:
 
 \begin{equation}
     x_{t-1}  = \begin{cases}
                f(q_\theta(y_{i-1}|y)) + p_\theta(x_{i-1}|x_i, c) - f(p_\theta(x_{i-1}|x_i, c)),\\ 
                \text{if t} > \text{ILVR stopping step}\\
                p_\theta(x_{i-1}|x_i, c), \quad\ \text{else} 
                \end{cases}
     \label{eq17}
 \end{equation}
 
 In \href{https://alonlevko.github.io/ilvr-tts-diff}{the project's demo website}\footnote{https://alonlevko.github.io/ilvr-tts-diff} we present samples obtained using different stopping steps.
 
 When the scaling parameters $N_{\rm F}$ and $N_{\rm T}$ are small, it is less likely that the reference spectrograms $y$ and sample spectrograms $x_0$ will comply with the ILVR condition in Eq.\ref{eq11}. Thus, picking the correct values of $N_{\rm F}$ and $N_{\rm T}$ is important for the voice conversion task and for balancing the old condition $c$ with our new condition. 
 
\begin{figure}[hb]
\centering
\begin{overpic}[width=0.9\linewidth]{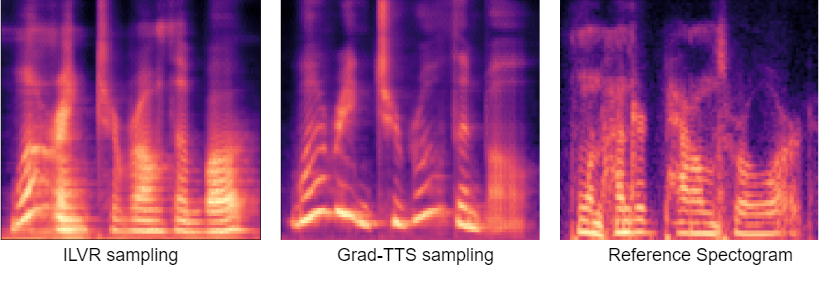}
\end{overpic}
\caption{Difference between Grad-TTS sampling and ILVR sampling for the same sentence. Low frequencies that are clear in the reference spectrograms, but absent in Grad-TTS sample appear in the ILVR sample.}
\label{fig3}
\end{figure}

\begin{figure}[h]
 \vspace{-0.3cm}
\caption{Example of spectrograms after low-pass filter for different scale factors. $N_{\rm F}=1$ for all spectrograms.}
\centering
\begin{overpic}[width=\linewidth]{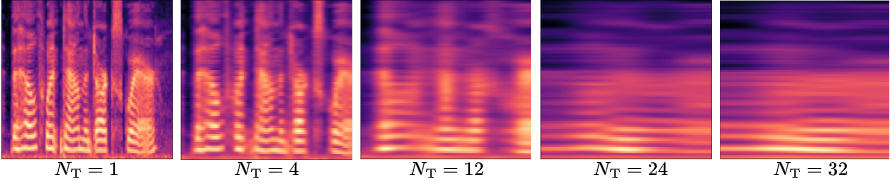}
\end{overpic}
\label{fig2}
 \vspace{-0.3cm}
\end{figure}

 The effects of changing the scale factors $N_{\rm F}, N_{\rm T}$ on the output of the filter can be seen in Figure \ref{fig2}. Since the filter is applied to spectrograms, it preserves the main frequency domain, which are, for speech, the F0 ferments, which allows us to use ILVR sampling to apply the features of the reference audio to the spectrograms generated. An example of the transfer of speech features can be seen in Fig.~\ref{fig3}.
 
 Our method is summarized in Alg.~\ref{grad-tts-ilvr}. In line 3, the algorithm first samples noise spectrograms from the prior distribution defined by the text encoder. In lines 4 to 19, the algorithm generates a sample from this noise iteratively, for $T$ steps. Lines 5 to 9 correspond to solving the SDE from Eq.\ref{eq6} using the Grad-TTS method. In line 10, the algorithm checks whether it needs to perform ILVR sampling or should continue with DDPM sampling from Eq.\ref{eq17}. If ILVR is to be performed, lines 11 to 15 apply the forward diffusion process $q_\theta$ to the reference spectrograms $y$. Line 16 is the ILVR step from Eq.\ref{eq16}. 
 
\begin{algorithm}[t]
 \caption{Grad-TTS with ILVR sampling}
 \label{grad-tts-ilvr}
\begin{algorithmic}[1]
  \STATE $\mu$: input text prior, $\tau$: temperature, $\beta_t$: noise schedule, $T$: time horizon, $j$: speaker index.
  \STATE $p_\theta$: unconditional DDPM, $f_{N_{\rm F}, N_{\rm T}}$: Filter, $s$: ILVR stoping step, $y$: reference data
  \STATE $x_T\sim \mathcal{N}(\mu, \tau^{-1}I)$
  \FOR{$i=T$ {\bfseries to} $1$}
  \STATE $t \leftarrow \frac{i}{N}$
  \STATE $z_t \leftarrow \beta_0  + (\beta_T - \beta_0)t$
  \STATE $dx_{i} \leftarrow \frac{1}{2}(\mu - x_{i} - p_\theta(x_{i}, \mu, t, j))$ 
  \STATE $dx_{i} \leftarrow \frac{dx_{i} z_t}{T}$
  \STATE $x_{i-1}' = x_{i} - dx_{i}$
      \IF{$i > s$}
        \STATE $n \leftarrow t\beta_0 + \frac{1}{2}(\beta_T - \beta_0)t^2$
        \STATE $m \leftarrow y \exp(\frac{-n}{2}) + \mu (1 - \exp(\frac{-n}{2}))$
        \STATE $v \leftarrow 1 - \exp(-n)$
        \STATE $\alpha \sim \mathcal{N}(0, \mathbf{I})$
        \STATE $y_{i-1} \leftarrow m + \alpha \sqrt{v}$
        \STATE $x_{i-1} \leftarrow f_{N_{\rm F}, N_{\rm T}}(y_{i-1}) + x_{i-1}' - f_{N_1, N_2}(x_{i-1}')$
      \ELSE
        \STATE $x_{i-1} \leftarrow x_{i-1}'$
      \ENDIF
  \ENDFOR 
  \STATE \textbf{return} $x_0$
\end{algorithmic}
\end{algorithm}

 \noindent{\bf Voice Conversion\quad}
 Picking the correct values of $N_{\rm F}$, $N_{\rm T}$, and the ILVR stopping step results in full voice conversion. This does not require any training, as the Grad-TTS model is already pre-trained and our method is only applied to sampling.
 
 To achieve the best voice conversion results, we experimented with different reference audio, text, and generated speakers. Based on the results, we chose these parameters as the voice conversion parameters of our method:  
 
 \begin{equation}
     \text{ILVR stop step} = 6 \quad\ 
     N_1 = 1 \quad\ 
     N_2 = 18
     \label{eq18}
 \end{equation}
 
 Note that picking small values for $N_{\rm F}$, $N_{\rm T}$ (For example $N_{\rm F} = 1$, $N_{\rm T} = 4$) will cause $f_{N_{\rm F},N_{\rm T}}(y_{t-1})$ to be similar to the reference spectrogram. Combining this with an ILVR stopping step that is too late will generate full voice conversion, but will ignore the required text. This causes the Grad-TTS prior spectrogram to be overrun by features from the reference spectrogram. In this case, the model will restore the reference audio with its original text correctly.  
 
 If $N_{\rm F}$ and $N_{\rm T}$, are too large (for example, $N_{\rm F}=1, N_{\rm T}=30$) ILVR sampling will only capture general features of the reference spectrogram. In this case, text conditioning is retained, and the generated sample will have speech features that are similar to the reference, but without full voice conversion. Samples for different values of $N_{\rm F}, N_{\rm T}$ can be found on \href{https://alonlevko.github.io/ilvr-tts-diff}{the project's demo website}. An example of the effect of different $N_{\rm T}$ values on spectrograms output by the filter can be seen in Fig.~\ref{fig2}.
 
{Picking $N_{\rm F} > 1$ results in the filter merging values for multiple frequencies together. We noticed that this has a negative effect on the quality of the output audio and it harms the output's similarity to the reference speaker. }
 
 \noindent{\bf Reference Audio that is not speech\quad}  The reference audio does not need to be the speech of a different person. In our method, ILVR sampling is performed in the spectrogram space. This means that the transition happens in the frequency domain and the generated audio is close to the audio of the reference in this domain. The user can pick music or a tone and create speech whose pitch and frequency are close to that of the tone. We noticed that this works best when the chosen reference is a low or high tone and that the pitch of the generated audio matched the pitch picked in the reference. Examples of this can be found at \href{https://alonlevko.github.io/ilvr-tts-diff}{the project's demo website}.
%  \footnotemark[2]
 
\section{Experiments}

\begin{table}[t]
\centering
\caption{MCD and MAE F0 Scores (Mean; lower is better)}
\label{tab:mcdsingle}
\begin{tabular}{lccc}
\toprule
Method    			                   & MCD  ($\downarrow$)  & MAE F0 ($\downarrow$) &Trained \\
\midrule

VoiceGrad \cite{kameoka2020voicegrad}  & 7.55         &--           & Yes  \\
TGAVC \cite{tang2021tgavc}             & 8.24         &--            & Yes \\
AutoVC \cite{qian2019autovc}           & 9.93         &--            & Yes \\
\midrule
DGC-VECTOR \cite{xiao2022dgc}          & 11.46        & 20.5            & No  \\
Our       			                   & 5.98         & 0.85            & No  \\
\bottomrule
\end{tabular}
%\end{table}
\medskip
% \begin{table}[htb]

%\end{table}
%\begin{table}[htb]
\medskip
\centering
\captionsetup{justification=centering}
\caption{MOS similarity and naturalness Scores}%\\ (Mean; higher is better)}
\label{tab:mossingle}
\begin{tabular}{lccc}
\toprule
Method    			                   & MOS ($\uparrow$)    & MOS-SIM ($\uparrow$) & Trained \\
\midrule

VoiceGrad \cite{kameoka2020voicegrad}  & ~3.2   & ~3.5     & Yes \\
TGAVC \cite{tang2021tgavc}             & 3.64   & 3.84     & Yes \\
AutoVC \cite{qian2019autovc}           & 3.25   & 3.08     & Yes \\
StarGANv2-VC \cite{li2021starganv2}    & 4.09   & 3.86     & Yes \\
\midrule
Ground Truth                           & 4.31   & 4.26     &--   \\
Baseline Grad-TTS                      & 3.8    &--        &--   \\
DGC-VECTOR \cite{xiao2022dgc}          & 4.04   & 2.98     & No  \\
Our       			                   & 3.6    & 3.96     & No  \\
\bottomrule
\end{tabular}
\medskip
\vspace{-0.5cm}
\end{table}

 In these experiments, we did not perform any training on a model. We used the pre-trained Grad-TTS model, which was trained on the Libri-TTS \cite{zen2019libritts} dataset. This model was trained on 247 speakers from the dataset. With it, a pre-trained HiFiGAN vocoder that was trained on the same dataset was used. 

\noindent{\bf Objective Evaluation\quad}
 Since this is a zero-shot voice conversion scenario, we used 8 speakers (4 Male and 4 Female) from the VCC2020 dataset \cite{yivoice}, which Grad-TTS had not been trained on. From each speaker, we randomly picked one utterance to be the reference. 
 
 For each speaker, we generated 247 samples with the same text sentence. Overall we evaluated 1976 samples. We used Dynamic time-warping Mel-cepstral distortion (DTW-MCD) \cite{kubichek1993mel} and mean absolute error (MAE) of F0 as the objective metric, where lower is better. We used the generated utterances as predictions and the reference utterances from VCC2020 as the ground truth data for calculating the metrics.
 
 {As can be seen in Tab.~\ref{tab:mcdsingle}}, in terms of MCD, the proposed method shows a sizable improvement over any other method known to us. Our method's MAE of F0 is significantly lower than for the other methods, {including the untrained DGC-VECTOR method,} and the frequencies are extremely close to the references. 

\noindent{\bf Subjective Evaluation\quad}  For the subjective evaluation, we used the mean opinion score (MOS) test. We used four Reference speakers from the VCC2020 dataset and randomly selected two utterances sampled from each. For control, we generated six samples of random speakers from the Libri-TTS dataset and we report their naturalness. The raters evaluated the naturalness and speaker similarity of utterances with scores between 1 and 5, where higher is better. 
 
 As can be seen in Tab.~\ref{tab:mossingle}, our proposed method achieved a MOS-Similarity score of $3.96$, which is a major improvement over the $2.98$ of DGC-VECTOR, which is a state of the art untrained voice conversion method. Regarding trained methods, the score of our method puts us at the higher end of state-of-the-art scores, with the best score being $3.89$ for StarGANv2-VC. In terms of MOS-Naturalness, our method achieves an MOS of $3.6$, whereas the MOS of the baseline Grad-TTS method is $3.8$, indicating a slight deterioration in terms of naturalness in comparison to the model we modified. This affects the similarity score, as human listeners find it more difficult to rate similarity between natural and less-natural speech utterances. 
 
\section{Conclusions}
By combining a voice DDPM with a spectrogram-domain conditioning method, we are able to perform TTS with voices unseen during training. This does not require any further tuning of the voice model or any auxiliary networks. Empirically, in the objective evaluation we demonstrate improvement in MCD and MAE F0 scores compared to previous methods. {In the subjective evaluation also demonstrate an improvement over other state of the art zero shot voice conversion method and comparable results to state of the art trained voice conversion method.}

\section{Acknowledgements}
This project has received funding from the European Research Council (ERC) under the European Unions Horizon 2020 research and innovation programme (grant ERC CoG 725974).

\bibliographystyle{IEEEtran}

\bibliography{mybib}

\end{document}